\documentstyle{l-aa}

\def\ccf{cross--correlation}

\def\secondip{\hbox{\rlap{\hbox{.}}\hbox{$''$}}}
\def\kms{\mbox{km~s$^{-1}$}}

\def\f{\mbox{$\ell$}}
\def\Ltot{\mbox{$L_A$}}
\def\Mtot{\mbox{$m_A$}}
\def\msun{\mbox{M$_\odot$}}
\def\s{\mbox{$\sigma$}}
\def\sw{\mbox{$\sigma_w$}}
\def\scc{\mbox{$\sigma_{cc}$}}
\def\velin{\mbox{$\langle v^2_r\rangle_\infty^{{1\over2}}$}}
\def\vel{\mbox{$\langle v^2_r\rangle^{{1\over2}}$}}
\def\vrm{\mbox{$\langle v_r\rangle$}}
\def\vrad{\mbox{$v_{r}$}}

\def\mlim{\mbox{$M_{lim}$}}
\def\mincir{\ \raise-2.truept\hbox{\rlap{\hbox{$\sim$}}\raise5.truept
\hbox{$<$}\ }}
\def\magcir{\ \raise-2.truept\hbox{\rlap{\hbox{$\sim$}}\raise5.truept
\hbox{$>$}\ }}

\def\etal{{\em et al.\/} }

\def\ie{{\em i.e.\/}}
\def\eg{{\em e.g.\/}}

\newcommand{\AJ}[3]{ #1, AJ #2, #3}
\newcommand{\AeA}[3]{ #1, A\&A #2, #3}
\newcommand{\AAS}[3]{ #1, A\&AS #2, #3}

\newcommand{\ApJ}[3]{ #1, ApJ #2, #3}

\newcommand{\MSAIt}[2]{ Mem. Soc. Astron. Ital. #1, #2}

\begin{document}
\thesaurus{ 05(10.07.2; 10.07.03 NGC 7099; 10.07.03 NGC 7078) }
\title{High resolution kinematics of galactic globular clusters. II.
On the significance of velocity dispersion measurements}
\author{Simone R. Zaggia \and Massimo Capaccioli \and Giampaolo Piotto}
\offprints{Simone R. Zaggia}
\institute{Dipartimento di Astronomia, Universit\`a di Padova,
           Vicolo dell'Osservatorio 5, I--35122 Padova, Italy.}
\date{Received November, 1992 ; accepted April, 1993}
\maketitle
\markboth{S.R. Zaggia, M. Capaccioli and G. Piotto: On the significance
of velocity dispersion measurements in globular clusters.}{}
\begin{abstract}
Small number statistics may heavily affect the structure of the broadening
function in integrated spectra of galactic globular cluster centers.
As a consequence, it is {\it a priori\/} unknown how closely line
broadening measurements gauge the {\it intrinsic\/} velocity dispersions at
the cores of these stellar systems.
We have tackled this general problem by means of Monte Carlo simulations.
An examination of the mode and the frequency distribution of the {\it
measured} values of the simulations indicates that the low value measured
for the velocity dispersion of M30 (Zaggia \etal\ 1992b) is likely a
reliable estimate of the  velocity dispersion at the center of this
cluster.
The same methodology applied to the case of M15 suggests that the steep
inward rise of the velocity dispersion found by Peterson, Seitzer and
Cudworth (1989) is real, although less pronounced.
Large-aperture observations are less sensitive to statistical fluctuations,
but are unable to detect strong variations in the dispersion wich occur
within the aperture itself.
\end{abstract}

\section{Introduction}
Since the pioneering work of Gunn and Griffin (1979), much effort has been
devoted to investigate the internal dynamics of galactic globular clusters
(GGCs), both theoretically and observationally.
The strategy is to derive the fundamental parameters of a GGC, such as mass and
mass--to--light ratio, by constraining models through the radial luminosity and
velocity dispersion profiles.
The latter are measured in two ways:
\begin{description}
\item{\it i)} either from the radial velocities of individual cluster stars,
or
\item{\it ii)} from the line broadening of the spectrum integrated over some
central area.
\end{description}
\noindent
The second approach is more economical than the first one, but it suffers
from the bias that the integrated spectrum is dominated by few bright
stars only, whose light overwhelms that of all the others within the
spectrograph aperture.
For this reason it seems important to investigate the statistical
significance of the measured values \s\ of the line broadening and their
correspondence to the desired estimate of the velocity dispersion \vel\
(see eq.~2).

The case of M15 is emblematic.
Peterson, Seitzer, and Cudworth (1989, hereafter PSC)  have reported that
the velocity dispersion at the very center of this core--collapsed (King
and Djorgovski 1984) GGC is as large as $25\pm7$ \kms, a fact implying a
steep inward rise of \vel.
This figure results from their interpretation of the measurements in
the framework of the small number statistics:
in fact, their sampling of \vel\ in the $3''\times 3''$ central area, made
using a circular aperture with $2R=1\secondip2$, produces values of \scc\
ranging from 8 to 30 \kms.
(Hereafter the subscript {\it cc\/} attached to \s\ indicates a measure of
\s\ obtained with the \ccf\ method; {\it cf.\/} Tonry and Davis 1979.)
Using a much larger aperture ($6''\times6''$) Dubath, Meylan, and Mayor
(1992, hereafter DMM) measure {\it directly\/} a lower value
$\scc=14.0\pm0.3$ \kms.
This second result, based on a larger number of stars, is statistically
more significant, but its poorer spatial resolution may likely wipe out a
rapid increase of \vel\ in the inner 1--2 arc seconds, if any.
Then the question is: how significant is the peak value
of \scc\ taken by PSC as the central value of \vel\ in M15\,?

In order to shed some light onto the general problem of the statistical
significance of the velocity dispersion measurements from integrated
spectra of GGCs, we have investigated the effects of the various parameters
on simulated spectra of the central regions of selected clusters.
In the next Section we present the methodology, and in Sect.~3 we discuss
its application to the case of M30, whose central velocity dispersion has
been measured by Zaggia \etal\ (1992b = Paper~I).
The case of M15 is discussed in Sect.~4.
Previous account of this matter has been given by Zaggia \etal\ (1992a).
\vfill

\section{Methodology for the simulations}
The width \sw\ of the broadening function of an integrated spectrum
represents the luminosity--weighted dispersion
\begin{equation}
\sw=\left[\frac{\sum_{i=1}^{N} \f_i\times(\vrad_i)^2}{\sum_{i=1}^{N} \f_i}
\right]^{1/2}
\end{equation}
of the radial velocities $\vrad_i$ of the stars with luminosity $\f_i$,
falling inside the area viewed by the spectrograph slit.
(Other effects such as seeing convolution of the spectrograph aperture,
convolution of the broadening function by stellar and cluster rotation, and
presence of close binary systems are ignored).
Thus \sw\ may differ from the unweighted radial velocity dispersion
for the totality $N$ of the stars within the aperture,
\begin{equation}
\vel=\left[\frac{\sum_{i=1}^{N}(\vrad {_i})^2}{N-1}\right]^{1/2}
\end{equation}
which is the quantity required in dynamical calculations.
In fact, the broadening function is dominated by the brightest stars.
Since they occur in a small number, the dispersion of their radial velocities
does not necessarily represent the {\it average \/} velocity dispersion within
the sampled area.

\begin{figure}
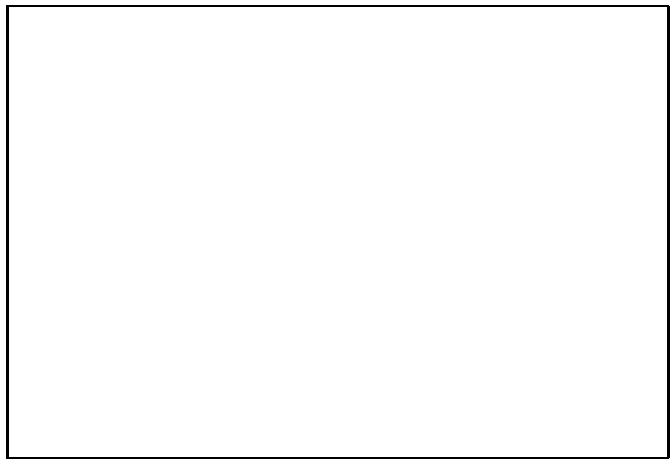
\label{LF}
\picplace{6cm}
\caption[]{The integrated luminosity function used in the simulations (solid
line) gives the number $\Phi(M_V)$ of stars brighter than $M_V$ (per magnitude
unit). The corresponding luminosity relative to the total luminosity (in
percent) is also shown (dashed line). \mlim\ values adopted for M15 and M30 are
plotted as upright dotted lines.}
\end{figure}

We have tested the effect of the small number statistics and of the star
luminosity--weighting in determining the value of \s, by means of
simulated spectra with {\it a priori\/} known \velin; this is the center of the
Gaussian distribution for \vel\ produced by poor statistics, \ie\ the
asymptotic value provided by eq.~2 with $N\rightarrow\infty$.
Our Monte Carlo approach accounts also for the effects of the measuring
technique used in Paper~I, which is based on a fit of the peak of the
\ccf\ function.
Each experiment consists of the following steps.
\begin{enumerate}
\item First we generate a random set of stellar luminosity values $\f_i$
with probability distribution given by a luminosity function (LF) appropriate
to metal poor GGCs (Ferraro 1990; see Fig.~\ref{LF}).
The set is closed when a pre--determined total luminosity $\sum\f_i=\Ltot$
is reached.
\item An integrated spectrum is then simulated by summing up, for each star
of the above set, a same (standard) spectrum weighted by the star
luminosity $\f_i$, and shifted in wavelength by an amount corresponding to a
random radial velocity with Gaussian probability distribution of fixed variance
\velin.
\item Photon and read--out noise are added to the synthetic cumulative
spectrum; we decided to tailor these effects to the observations of Paper~I
(global efficiency of the 3.6 m\,+\,CASPEC, and characteristics of the CCD
used there).
\item Finally, the broadening \scc\ of the synthetic spectrum is measured
by the same \ccf\ technique adopted for the real spectra.
\end{enumerate}
\noindent
Input parameters for a set of Monte Carlo simulations are:
\begin{description}
\item{---} the total luminosity $\Ltot=\sum\f_i$ within the slit aperture A;
\item{---} the cut--off value \mlim\ for the LF: stars fainter that \mlim\
are ignored in building up the set at step~1.
Thus \mlim\ is the fixed value which controls the number of stars actually
concurring to the formation of the simulated spectrum;
\item{---} the intrinsic (unweighted) radial velocity dispersion \velin.
\end{description}
\noindent
For all the experiments we made use of a high signal--to--noise  spectrum of
HD~122563, the template star of Paper~I, obtained with CASPEC at the ESO 3.6~m
telescope (resolution of ${\rm FWHM}=0.156$\,\AA, equivalent to 9.0 \kms\
at $\lambda=5200$\,\AA).
We have verified that the results do not change by utilizing a lower
signal--to--noise spectrum of a typical GGC star.
We have also tested the use of LFs other than the one adopted here
(Fig.~\ref{LF}), finding no appreciable difference even with LFs of metal
rich GGCs (\eg\ NGC 2808; Ferraro \etal\ 1990).
Finally, simple considerations show that no major consequence is expected in
the results by the fact that stars are assumed to have all the same mass and so
the same velocity dispersion, provided that \mlim\ is faint enough.
In fact the difference between the mass of a turn--off star and one at 3.5 mag
below it, \ie\ our typical cut--off, is approximately 0.2 \msun.
With the assumption of isothermal velocity distribution, this difference
corresponds to an increase of $\simeq 20\%$ of the velocity dispersion with
respect to that of the turn--off stars.

\begin{figure*}\label{m30:histo}
\picplace{13.5cm}
\caption[]{The shaded area in each panel is the frequency distribution of the
{\it measured\/} line broadening \scc\ from 1000 Monte Carlo experiments
with the indicated value of the input velocity dispersion \velin\ (arrow)
and fixed $\mlim=+3.5$ mag.
The solid line is the distribution of the luminosity--weighted dispersion
\sw\ for the input set of stars (see eq.~(1)).
The upper left panel reports also the histogram of \vel\ (dotted line) for
that particular set of experiments.}
\end{figure*}

\section{The meaning of \scc: an application to M30}
Here we show how small number statistics and gauging of the broadening function
operate by reporting on Monte Carlo experiments tailored to our observations of
M30 (Paper~I).
This is a GGC which, by analogy with M15, is suspected to possess a central
spike in the velocity dispersion (Zaggia \etal\ 1991).
In all simulations \Ltot\ has been fixed to $-3.5$ mag to match the total light
($\Mtot=11.0$ mag) within the centered aperture of $4\secondip6\times6''$ used
for the spectra of Paper~I.
Simulations have been made for three values of $\mlim=1.5$, 3.5, and 7.5
mag, and four values of the input velocity dispersion $\velin=6$, 12, 18, and
24 \kms.

The results for groups of 1000 simulations are shown in the panels of
Fig.~\ref{m30:histo} for the four values of \velin\ (indicated by an arrow
in the figures), keeping \mlim\ fixed at 3.5 mag.
The shaded area marks the distribution of the output values \scc\ of the
\ccf\ method applied to the simulated spectra.
We also plot the luminosity--weighted velocity dispersion \sw\ of the
input sets of stars (solid line), as it results from eq.~(1).

The effect of the small number statistics stands out clearly in the upper-left
plot of Fig.~\ref{m30:histo}, where we have also plotted the distribution of
\vel\ (dotted line); this unweighted equivalent of \sw\ is calculated for
each set of stars with eq.~(2).
The histogram of \vel\ (truncated in its upper part) has a smaller dispersion
than those of \sw\ and \scc.
The mean of the distribution of \vel\ is equal to the input value \velin,
and the scatter is due to the small number of stars:
this implies that the standard deviation of \vel\ is Poissonian.
In fact, with a mean number of $\sim74$ stars per set (brighter than \mlim)
and for $\velin=6$ \kms, the Poissonian standard deviation is 0.49 \kms,
well in agreement with the standard deviation of the distribution of \vel\
derived from our simulations (0.51 \kms).

\begin{figure}
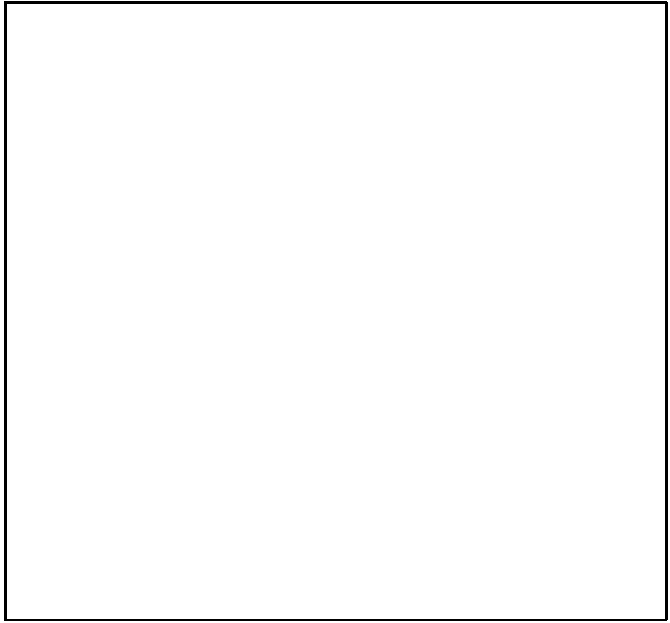
\label{m30:inout11}
\picplace{8.2cm}
\caption[]{Comparison between input (\vel) and output (\scc) values of the
velocity dispersion from Monte Carlo simulations tailored to our (Paper I)
observations of M30.
The horizontal dotted line represents the measure of \scc\ reported in
Paper~I.}
\end{figure}

The luminosity weighting permits some bright stars or group of stars to
dominate the velocity dispersion.
So, the distribution of \sw, calculated with eq.~(1), comes out broader
than that of \vel.
For the experiment with $\velin=6$ \kms, the distribution of \sw\ has a
mean value of 5.5 \kms\ and a standard deviation of 1.1 \kms.
This behavior is confirmed by the other simulations, and could be regarded as
the limit to our kinematical measurements: in fact, what we measure in a
spectrum is \sw, which is most probably $\le\vel$.
For instance, in the application to the case of M30 we find that the most
probable value of \sw\ is lower, by $\sim10\%$, than the input \velin,
and that the associated uncertainty (at 1 sigma level) is $20\%$ of
$\langle\sw\rangle$.
While the cut-off of the LF moves towards fainter magnitudes,
the intrinsic error lowers.
This is the reason why we have chosen $\mlim=7.5$ mag;
the integrated light of all stars brighter than this limit is 95\% of the total
light encircled by the spectrograph aperture.

By looking at the successive panels of Fig.~\ref{m30:histo} it is apparent
that the distributions of \scc\ and of \sw\ differ more and more as
\velin\ increases.
The distribution of \scc\ is also increasingly skewed towards the low
values.
This is due to the somewhat unsmooth form of the \ccf\ function.
In fact, with a high value of \vel, the \ccf\ function become multimodal,
with peaks produced by bright single stars or groups of stars with radial
velocity far from the mean.
All but one of these peaks are ignored by the fitting algorithm, which fits
the highest peak only;
consequently the measured dispersion is systematically lower than the
input value.
Paradoxically, this effect increases with the spectral resolution.

\begin{figure}\label{m15:vel}
\picplace{6.5cm}
\caption[]{Mean radial velocity \vrm\ for each of the PSC spectra of M15,
scaled to the average radial velocity of 120 single cluster stars (systemic
velocity), plotted against the corresponding velocity dispersion \scc\ (large
solid dots).
Error bars are from PSC.
The small dots show the behaviour of 1000 simulations made with $\velin=20$
\kms.
There is no clear correlation between \scc\ and \vrm.
This fact suggests that the largest values of \scc\ are not likely due to
anomalously large radial velocities of dominating bright stars.}
\end{figure}

The difference between the input and output values of the velocity
dispersion in these simulations of M30 can be better appreciated in
Fig.~\ref{m30:inout11}, where the mode of the distribution of \scc,
together with its half width at half maximum, is plotted against the
corresponding input value of \vel\ for three values of \mlim.
It is clear that, with the instrumental set up of Paper~I, any result
giving $\scc\magcir 10$ \kms\ would underestimate, {\it on average\/}, the
{\it intrinsic \/} velocity dispersion of M30.
In Paper~I we have measured $\scc=6.0\pm0.6$ \kms, well in the range where
$\scc\simeq\vel$.
Note however that, the same figure for \scc\ would be (marginally) in
accord also with values of \vel\ as high as $\simeq 25$ \kms\ or more.
We shall expand the discussion on this important point in the next Section.

It can be of some interest to study the effects of changing the size of the
integration aperture.
We have thus investigated the relation between \scc\ and \vel\ for an
aperture of $1''\times1''$ (about the same size of that used by PSC for
M15), which, at the center of M30, encircles a total luminosity \Ltot\ of
-0.7 mag (or $\Mtot=13.8$).
We have explored the parameter space using the same values of \mlim\ and
\vel\ as in the previous experiments, taking care to reject all the
experiments in which the total light is contributed by one star only.
As expected, the departure of the output from the input velocity dispersion
increases dramatically.
The measured \scc\ seems to reach a maximum between $\vel=10$ and 20 \kms,
then it decreases.
Consequently, under these conditions a measure of $\scc=5\div8$ \kms\ from a
{\it single\/} spectrum would have no other meaning but that of a possible
lower limit.

\begin{figure}
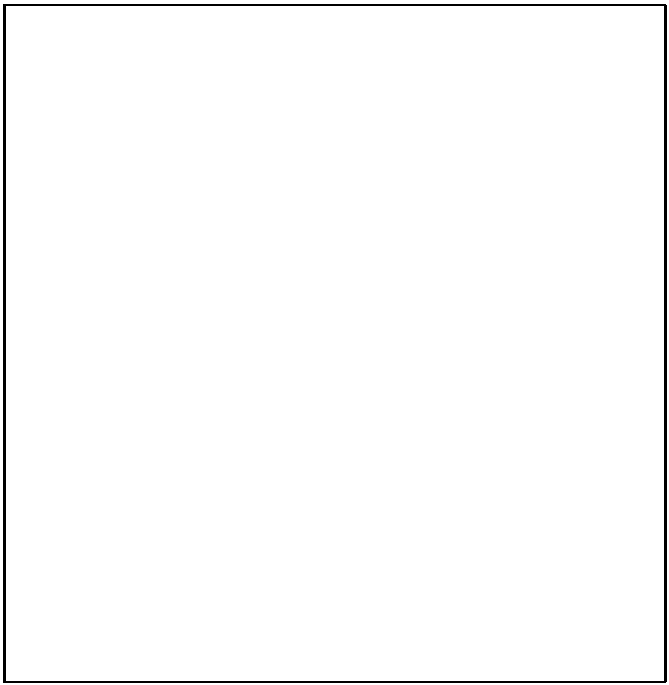
\label{m15:inout13}
\picplace{9cm}
\caption[]{Same as Fig.~\ref{m30:inout11}, but for M15 as observed by
PSC.
The data refer to the set of simulations with $\mlim=+6.8$ mag. The
horizontal dashed line is the mean of all the PSC values, while the dotted
line is the mean after rejection of two measurements deviating more than 3
sigma.}
\end{figure}

\section{The central \scc\ in M15}
The already mentioned claim by PSC of a steep gradient of the velocity
dispersion in M15, consequence of the high value of the central $\scc \ge
25$ \kms, seems contradicted by the measure of $\scc=14.0$ \kms\ reported by
DMM.
We shall note here that the most immediate sign of a statistical bias,
\ie\ the presence of a correlation of the velocity dispersion with the
corresponding absolute value of the excess cluster systemic velocity,
is not present in the PSC data (Fig.~\ref{m15:vel}).
To reconcile the two apparently conflicting observations by PSC and DMM, we
tailored the parameters of our Monte Carlo simulations machinery to M15.

In Fig.~\ref{m15:inout13} we report the results of the simulations for the
PSC parameter--set.
Here \velin\ varies from 5 to 40 \kms\ in steps of 5 \kms, $\mlim=6.8$ mag, and
$\Ltot=-1.7$ mag ($\Mtot=13.5$) as this is the (average) luminosity
collected by the circular aperture of $1\secondip2$ used by PSC to sample
the velocity dispersion in the core of M15.
What has been just discussed for M30 is still valid: $\scc<\velin$ for
$\velin > 10$ \kms.

\begin{figure}
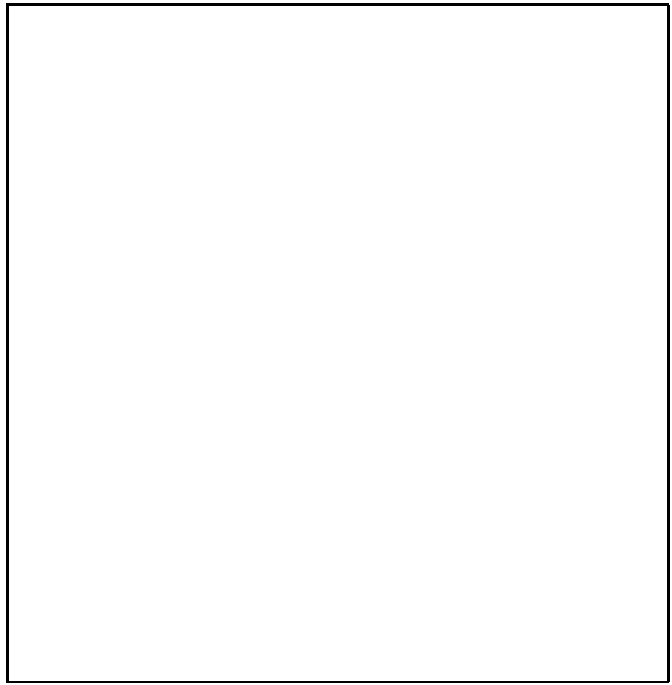
\label{m15:histo_pet}
\picplace{9cm}
\caption[]{The dotted shaded area in each panel represents the frequency
distribution of \scc\ for 1000 Monte Carlo experiments, tailored to the
PSC observations of M15, for the indicated value of the input velocity
dispersion \velin\ and fixed $\mlim=+6.8$ mag.
The solid shaded area is the histogram of the measurements of M15 velocity
dispersion by PSC.}
\end{figure}

For a better comparison with the PSC observations, in the various panels of
Fig.~\ref{m15:histo_pet} we plot the frequency distributions of \scc\ values
resulting from the 1000 simulations made for each input value of \velin.
Superimposed there is the histogram of the single measures of \scc\ by PSC
(their Table~4).
It is clear from the panels of Fig.~\ref{m15:histo_pet} that:
\begin{itemize}
\item Simulations allow to exclude that the central velocity dispersion of M15
can be as low as 10 \kms.
\item There are a chance to measure $\scc\ge20$ only if $\velin\magcir15$.
\item The peak of the values measured by PSC is nearly coincident with the
mode of the distributions of \scc\ only when $15<\velin<30$.
\end{itemize}
The PSC claim of $\vel\geq25$ \kms\ at the center of M15 rests on the
distribution of the results of their eight independent measurements of \scc\
made around the cluster center.
Six of them lie in the interval $8.4\leq\scc\leq11.8$ \kms\ (with a mean
value of 10.2 \kms), but the other two are of $23.6\pm3.1$ and $30.0\pm4.3$
\kms\ respectively, too large for $\velin\mincir15$ \kms\ according to our
simulations.
A more quantitative evaluation of the comparison between observations and
simulations has been made using the Kolmogorov--Smirnov (KS) test.
The KS probability of the match of PSC data with the Monte Carlo distributions
of Fig.~\ref{m15:histo_pet} is maximum (70\%) for $\velin=20$ \kms, but it is
still large at both $\velin=15$ and 25 \kms\ (63\% and 53\% respectively).

\begin{figure}
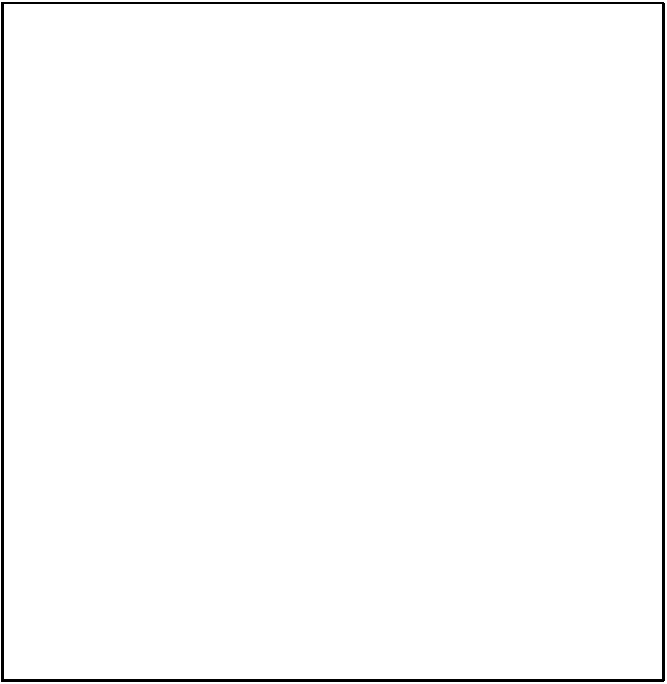
\label{m15:inout10}
\picplace{9cm}
\caption[]{Same as Fig.~\ref{m15:inout13}, but for a larger aperture of
$6''\times6''$ used for M15 by DMM.
Again the data refer to the $\mlim=+6.8$ simulation.
The horizontal dotted line represents the \scc\ of M15 measured by DMM.}
\end{figure}

In conclusion, our numerical experiments confirm that, within $R\sim
1\secondip5$ from the center of M15, the velocity dispersion is likely $\sim
20$ \kms\ (most probable value), in fair agreement with the qualitative
interpretation given by PSC to their observations:
``{\it We conclude that the low values for the central dispersion are those
dominated by one or two stars, and that the higher values are more
representative of the dispersion of the background light\/}'' (verbatim).
Indeed, the multiple sampling of \scc\ is the winning strategy to overcome
resolution and statistical problems at once.
Clearly, a single measure of \scc\ through an aperture as small as the one
adopted by PSC would be of no meaning.
At the same time, we shall see that the aperture used by DMM is large
enough to wipe out fine kinematical features.

\begin{figure}
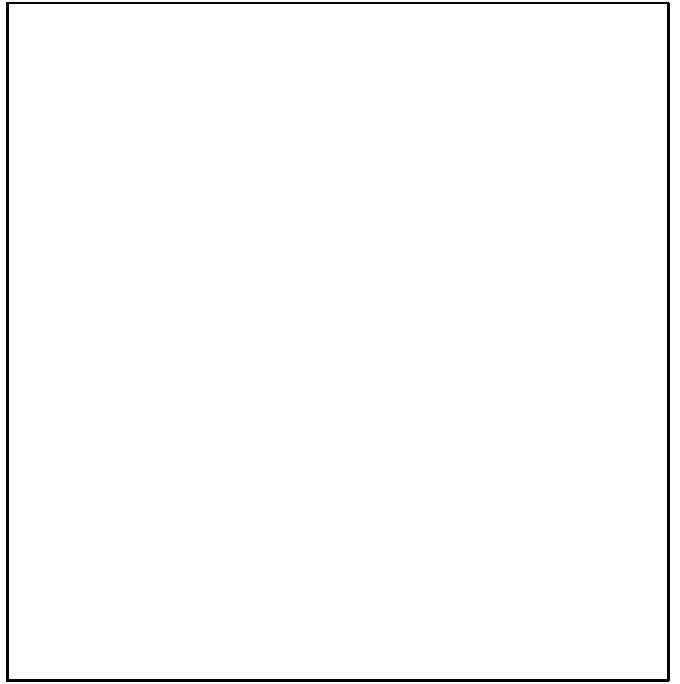
\label{m15:histo_mey}
\picplace{9cm}
\caption[]{Same as Fig.~\ref{m15:histo_pet}, with simulation parameters
tailored to the observations of M15 by DMM.
Their value of \scc\ is marked by an arrow.}
\end{figure}

DMM observed M15 with CASPEC through a centered aperture of $6''\times6''$
(as we did for M30).
The data plotted in Fig.~\ref{m15:inout10} have been computed with the same
set of parameters as in Fig.~\ref{m15:inout13} but $\Ltot=-4.5$ mag (or
$\Mtot=10.7$).
We see that DMM measure of $\scc=14.0$ \kms\ is compatible with a similar low
value for \velin, but it is consistent with values as large as $\velin=30$
\kms.
This is confirmed by the frequency distributions of the \scc\ values
for fixed \velin\ (Fig.~\ref{m15:histo_mey}).
When $\velin > 15$ \kms, the measure of DMM (arrow) is near the peak of the
distribution of \scc.

\section{Conclusions}
{}From the results of the simulations of the spectra of the central regions
of GGCs, we can draw the following conclusions:
\begin{enumerate}
\item For M30, our measure of $\scc=6.0\pm0.6$ \kms\ (Paper~I) is well in the
range where the simulations shows that $\scc\simeq\vel$.
\item We confirm {\it quantitatively\/} the results of the discussion by PSC
about the central velocity dispersion of M15.
\item The observations of M15 by DMM are consistent with an inward rise of
the radial velocity dispersion, although possibly not as steep as claimed by
PSC.
\end{enumerate}

\noindent
Finally, we find that a single integrated spectrum through a {\it large
enough\/} centered aperture is capable of providing just a lower value of
\vel.
The multiple independent sampling technique pioneered by Peterson \etal\
(1989) seems thus essential to investigate fine kinematical structures of
the cores of GGCs.

\end{document}